\documentclass{pnastwo}

\pdfoutput=1

\usepackage{float}
\usepackage{epsfig}
\usepackage{enumerate}
\usepackage{multirow}
\usepackage{array}
\usepackage{comment}

\usepackage{amssymb,amsfonts,amsmath,graphicx,multirow}

\newcommand{\me}{$M_{\oplus}$}
\newcommand{\re}{$R_{\oplus}$}

\newcommand{\icarus}{Icarus} 
\newcommand\apj{Astrophys.~J.~}                                          
\newcommand\apjl{Astrophys.~J. Lett.~}                                          
\newcommand\apjs{Astrophys.~J. Suppl.~}                                         
\newcommand\aap{Astron. Astrophys.~}                                            
\newcommand\mnras{Mon. Not. R.~Astron. Soc.~}

\newcommand\pasp{Pub.~Astron.~Soc.~Pacific~}                
\newcommand\nat{Nat.~}

\newcommand{\cN}[1]{\mathcal{N}}
\newcommand{\pn}[1]{\mbox{$(#1)$}}

\def\gsim{\;\rlap{\lower 2.5pt
 \hbox{$\sim$}}\raise 1.5pt\hbox{$>$}\;}
\def\lsim{\;\rlap{\lower 2.5pt
   \hbox{$\sim$}}\raise 1.5pt\hbox{$<$}\;}

\contributor{Submitted to Proceedings
of the National Academy of Sciences of the United States of America}
\url{www.pnas.org/cgi/doi/10.1073/pnas.0709640104}
\copyrightyear{2011}
\issuedate{Issue Date}
\volume{Volume}
\issuenumber{Issue Number}

\begin{document}


\title{The Structure of Exoplanets}

\author{
David~S.~Spiegel\affil{1}{Institute for Advanced Study, Princeton, NJ
  08540}\footnotetext{dave@ias.edu},
Jonathan~J.~Fortney\affil{2}{Department of Astronomy \& Astrophysics,
  University of California, Santa Cruz, CA 95064},
Christophe~Sotin\affil{3}{Jet Propulsion Laboratory, Caltech, 4800 Oak
  Grove Drive, Pasadena, CA 91109}}

\footlineauthor{Spiegel, Fortney, \& Sotin}

\contributor{Submitted to Proceedings of the National Academy of
  Sciences of the United States of America}

\maketitle

\begin{article}

\begin{abstract}
The hundreds of exoplanets that have been discovered in the past two
decades offer a new perspective on planetary structure.  Instead of
being the archetypal examples of planets, those of our Solar System
are merely possible outcomes of planetary system formation and
evolution, and conceivably not even terribly common outcomes (although
this remains an open question).  Here, we review the diverse range of
interior structures that are known to, and speculated to, exist in
exoplanetary systems --- from mostly degenerate objects that are more
than 10 times as massive as Jupiter, to intermediate-mass Neptune-like
objects with large cores and moderate hydrogen/helium envelopes, to
rocky objects with roughly the mass of the Earth.
\end{abstract}


\abbreviations{H/He, KOI, RV, Gyr}

\section{Introduction}
\label{sec:intro}
How can we, from many light years away, learn about the interior
structure of exoplanets?  Radial velocity (RV) observations provide
minimum masses of exoplanets; transit observations provide planet
radii.  Taken singly, neither is terribly informative about planet
structure.  However, when we know both the mass and the radius of a
planet we may learn much more about the interior structure.  The {\it
  Kepler} satellite \cite{borucki_et_al2011, batalha_et_al2013} is a
space-based, transit-detecting mission that, as of early 2013, has
identified of order $\sim$100 planets and roughly 3000 planet
candidates, of which the vast majority are almost certainly real
\cite{morton+johnson2011, fressin_et_al2013}.  Thanks to {\it Kepler}
and ground-based efforts such as the HAT and Super-WASP transit
surveys \cite{bakos_et_al2004, pollacco_et_al2006}, there are now more
than 200 known planets with measured masses and radii, spanning a
range of irradiation conditions.

\begin{figure*}[t]
\begin{center}
\includegraphics[width=8.5cm,angle=0,clip=true]{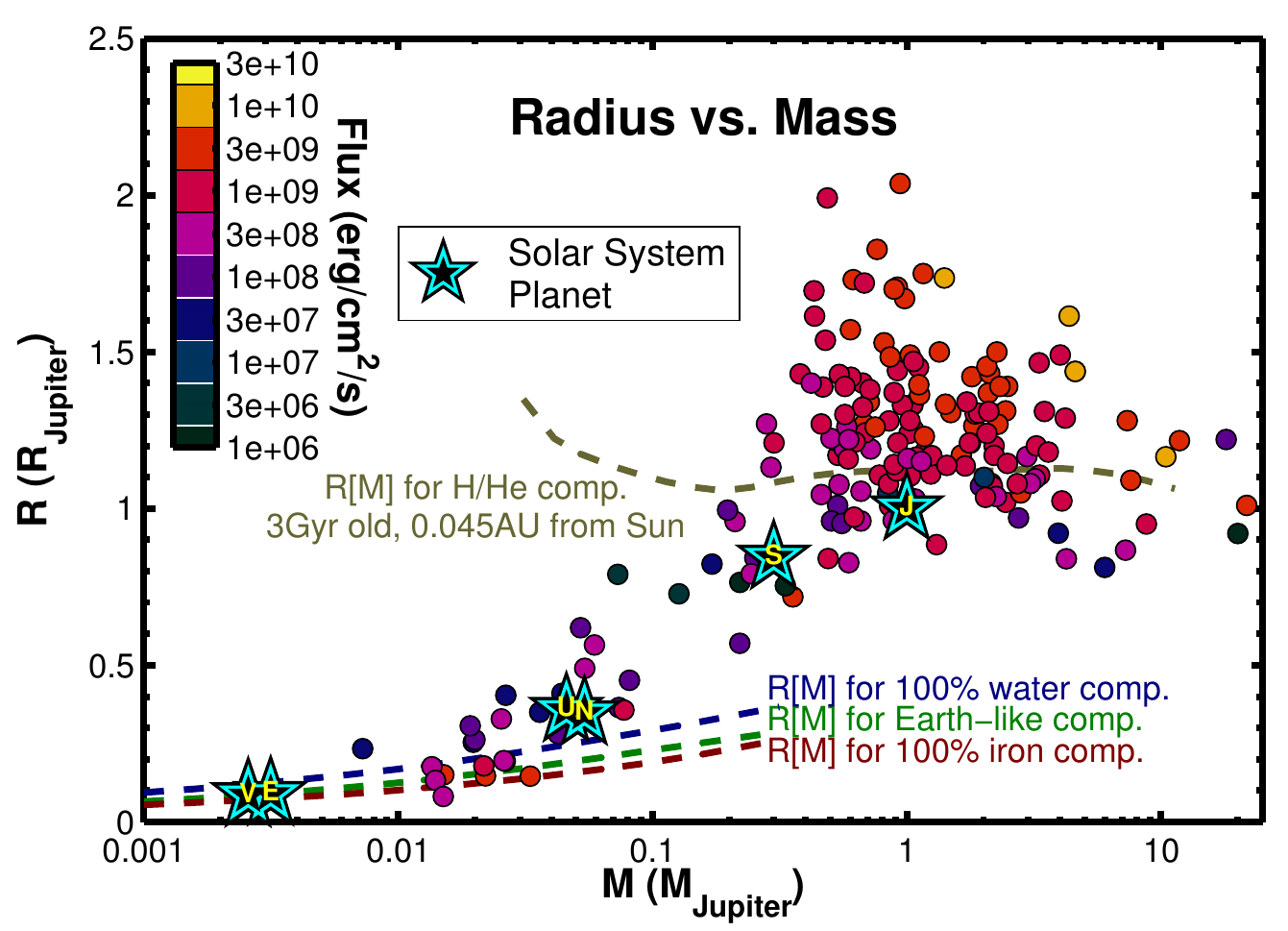}
\includegraphics[width=8.5cm,angle=0,clip=true]{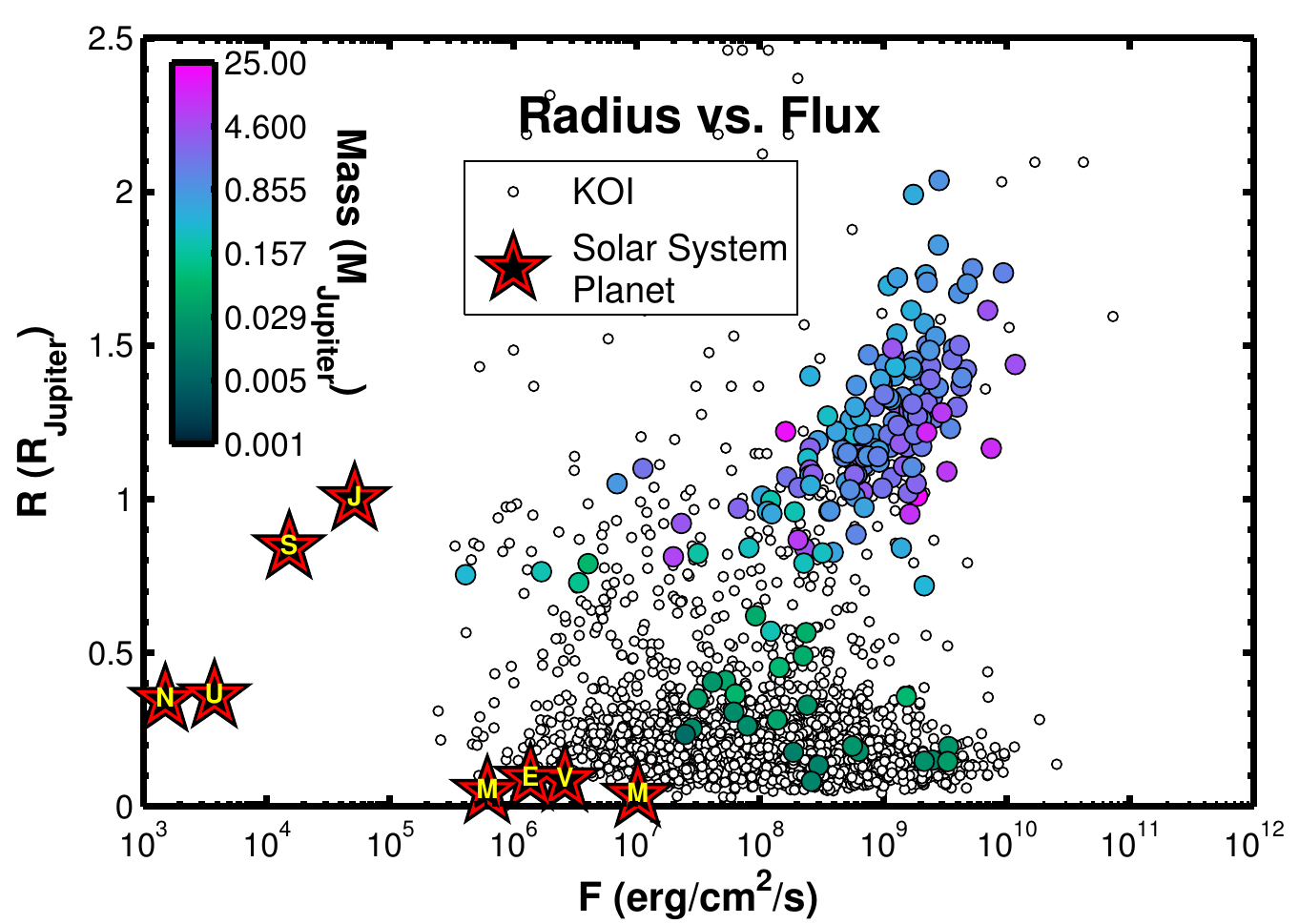}
\end{center}
\caption{\small Radius vs. mass (left panel) and vs. incident
  irradiating flux (right panel) for the confirmed exoplanets, the
  {\it Kepler} candidate planets ({\it Kepler} objects of interest ---
  KOI), and the planets of our Solar System.  The $\sim$200 confirmed
  planets in this figure are represented with filled large circles;
  KOI (in the right panel) are represented with small white circles;
  Solar System planets are represented with large pentagrams.  For
  comparison, the left panel shows the radius-vs.-mass relationship
  for Earth-like composition, which precisely matches the
  (mass,radius) values for Venus and Earth \cite{grasset_et_al2009}.
  Planets span masses from less than Earth's to tens of times
  Jupiter's, radii from less than Earth's to more than double
  Jupiter's, and incident fluxes from nearly 0 --- in the case
  HR~8799b, at $\sim$70~AU \cite{marois_et_al2008} --- to nearly
  $10^{12}$~erg~cm$^{-2}$~s$^{-1}$.  The planets with the largest
  radii tend to be close to Jupiter's mass and highly irradiated.}
\label{fig:Rad_vs_stuff}
\end{figure*}

Figure~\ref{fig:Rad_vs_stuff} portrays how planet radii relate to
masses and incident fluxes, among the known planets (including the
Solar System planets) and the {\it Kepler} candidates ({\it Kepler
  objects of interest} -- KOI) \cite{wright_et_al2011,
  schneider_et_al2011}.  Several trends are apparent in the data:
planets with the largest radii tend to be near Jupiter's mass and
highly irradiated; and, {\it Kepler} seems to find low-radius planets
at a wide range of orbital separations, including some small planets
that are extremely highly irradiated.

In the remainder of this paper, we discuss what is known of the
structure of the most massive planets (``Gas Giants''), of
intermediate-mass planets (``Neptunes''), and of low-mass planets
(``Terrestrial and Ocean Planets'').  We conclude by considering how
our knowledge of exoplanet structure might improve over course of the
next decade.

\section{Gas Giants}
\label{sec:gasgiants}
Jupiter and Saturn are essentially giant spheres of hydrogen and
helium (``H/He'') with smaller contributions from heavier elements and
complex molecules.  Many of the known exoplanets have similar mass and
similar radius to our local gas giants, and probably have roughly
similar bulk structure.  The atmosphere, or weather layer, of such an
object is a thin outer region that is of roughly the same relative
depth as the skin of a grapefruit, and is typically defined as the
region above the radiative-convective boundary,\footnote{The
  radiative-convective boundary is the region bounding the convective
  interior of a planet in which heat transport is dominated by
  convective eddies.  This boundary occurs essentially where the
  vertical temperature gradient becomes subadiabatic and therefore
  stable against convection.} which can be at pressures of order
$\sim$kilobar for the most strongly irradiated planets and which
occurs in the vicinity of $\sim$1~bar in gas giants subjected to lower
irradiation, such as Jupiter and Saturn.  Below the
radiative-convective boundary, there is a deep envelope extending
almost the entire radius of the planet in which opacities are high
enough that heat must be transported via convection; this region is
presumably well-mixed in chemical composition and specific entropy.
Some gas giants have heavy-element cores at their centers, although it
is not known whether all such planets have cores.

Gas-giant planets of roughly Jupiter's mass occupy a special region of
the mass/radius plane.  At low masses, liquid or rocky planetary
objects have roughly constant density, and suffer little compression
from the overlying material.  In such cases, $R_p \propto M_p^{1/3}$,
where $R_p$ and $M_p$ are the planet's radius mass.  At high masses,
for objects that have had time to cool, electron degeneracy pressure
becomes significant, and the mass/radius relation changes such that
the radius scales as $R_p \propto M_P^{-1/3}$.  Note that the ``H/He
comp.'' curve in Fig.~\ref{fig:Rad_vs_stuff} does not display this
behavior at low masses, because this curve is calculated for highly
irradiated objects that are a mere 0.045~AU from their Sun-like star,
which prevents them from reaching their ``zero-temperature radius'' in
3~Gyr (three billion years).  For cold spheres of H/He, it has long
been appreciated that the maximum in the mass/radius relation occurs
near four times Jupiter's mass \cite{zapolsky+salpeter1969,
  burrows_et_al2001}, with a broad peak at just over 1~$R_J$ (where
$R_J$ is Jupiter's radius), extending from $\sim$a tenth of Jupiter's
mass to $\sim$100$\times$ Jupiter's mass.  For a planet to be smaller
than the H/He minimum radius requires the presence of a significant
heavy-element component, in the form of a core in the object's center
or high-metallicity material well mixed throughout the envelope (where
``metals'' is taken to mean elements heavier than helium).  For a
planet to be larger than the zero-temperature radius ($\sim$1~$R_J$)
requires it to have a high enough temperature that pressure from ions
becomes a significant fraction of that due to electrons.

Prior to the discovery of transiting planets, it was assumed that
$\sim$billion-year-old planets --- even ones on close-in orbits around
their stars (so-called ``hot Jupiters,'' with orbits lasting $\sim$a
week or less) --- would have cooled and shrunk to near their
asymptotic radius \cite{guillot_et_al1996, saumon_et_al1996}.
However, the discovery of transiting hot Jupiters such as HD~209458b
\cite{charbonneau_et_al2000} with its $\gtrsim$1.3~$R_J$ radius,
suggested that some planets are a good deal puffier than expected
\cite{burrows_et_al2000}.  Many mechanisms have been suggested in the
literature to explain the apparently inflated radii of some of the
hottest hot Jupiters \cite{baraffe_et_al2010, fortney+nettelmann2010};
what these mechanisms all have in common is that they increase the
bulk entropy of the planet above what might have been expected from
naive cooling models.\footnote{There are some uncertainties in the
  proper equation of state for H/He/metals mixtures
  \cite{saumon_et_al1995, militzer+hubbard2013}, but the attendant
  uncertainties in radius are insufficient to explain the very large
  radii ($\gtrsim$1.5~$R_J$) occasionally seen.}  This increase can
come either from \pn{i} retarding the evolutionary cooling (via
enhanced atmospheric opacity or modified atmospheric thermal profiles
\cite{guillot+showman2002, burrows_et_al2007, perna_et_al2010b}), or
\pn{ii} from additional power added in the planet's deep interior (via
tidal \cite{bodenheimer_et_al2001, miller_et_al2009} or Ohmic
\cite{batygin+stevenson2010, wu+lithwick2013} dissipation).

It remains uncertain whether a single inflation mechanism
predominates.  For the most highly-inflated objects, however,
mechanisms of type \pn{i} might not work.  Close-in planets are
presumably tidally locked (although see \cite{arras+socrates2010}) and
have a permanent nightside through which they cool more efficiently
\cite{guillot+showman2002, budaj_et_al2012}; accounting for the
enhanced night-side cooling efficiency suggests that mechanisms that
merely reduce the cooling rate might not be able to explain the highly
inflated planets, and some power must be deposited in the deep
interior \cite{spiegel+burrows2013}.  Despite the lack of a consensus
mechanism, it is clear that objects must either be quite young
\cite{baraffe_et_al2003} or very highly irradiated
\cite{fortney_et_al2011b, weiss_et_al2013} (see
Fig.~\ref{fig:Rad_vs_stuff}) in order to have significantly inflated
radii.  The lowest irradiation experienced by any planet with a radius
more than 1.5$\times$ Jupiter's is the
$\sim$$10^9$~erg~cm$^{-2}$~s$^{-1}$ incident on Kepler-12b, which
still exceeds the solar flux upon Jupiter by a factor of $10^4$.

One of the key structural uncertainties for many of the known
gas-giant planets is whether they have cores in their centers.  (It is
also not known, at present, whether Jupiter has a heavy-element core
at its center, although Saturn must have one.)  The presence or
absence of cores is of interest both as it relates to our knowledge of
the the planets themselves and because it bears upon their formation
mechanism --- whether by a runaway process of accreting gas onto
$\sim$10-$M_\oplus$ cores \cite{mizuno_et_al1978} or via gravitational
instability of the protoplanetary gas disk \cite{boss2010}.
Unfortunately, it is essentially impossible to learn whether the
extremely inflated planets have cores (though, if they do, the larger
the core mass, the greater the additional power that is required to
explain their radii).  Some of the known transiting gas giants,
however, have smaller radii (at their known masses) than a H/He
composition can produce.  These planets must have a significant
heavy-element component.  The inferred metal fraction appears to be
correlated with the metallicity of the planet's host star
\cite{guillot_et_al2006, burrows_et_al2007}, suggesting that more
metal-rich protoplanetary environments lead to more metal-rich
planetary compositions, perhaps in the form of rocky cores.

Some exotic objects orbiting other stars do not have direct structural
analogs in our Solar System.  One planet that falls between our local
archetypcal categories is the enigmatic HD~149026b
\cite{sato_et_al2005}.  Although 20\% more massive than Saturn, its
radius is 22\% {\it smaller}, which suggests that its heavy-element
content is greater than the entire mass of metals (outside the Sun) in
the Solar System, in the range of $\sim$60--110-$M_\oplus$ of metals
\cite{fortney_et_al2006b, burrows_et_al2007}.  In this respect,
despite its greater-than-Saturn mass, this planet is perhaps more
similar in structure to Uranus- and Neptune-like planets, the subject
of the following section.

\section{Neptunes}
\label{sec:neptunes}
\begin{figure*}[t]
\begin{center}
\includegraphics[width=14cm,angle=0,clip=true]{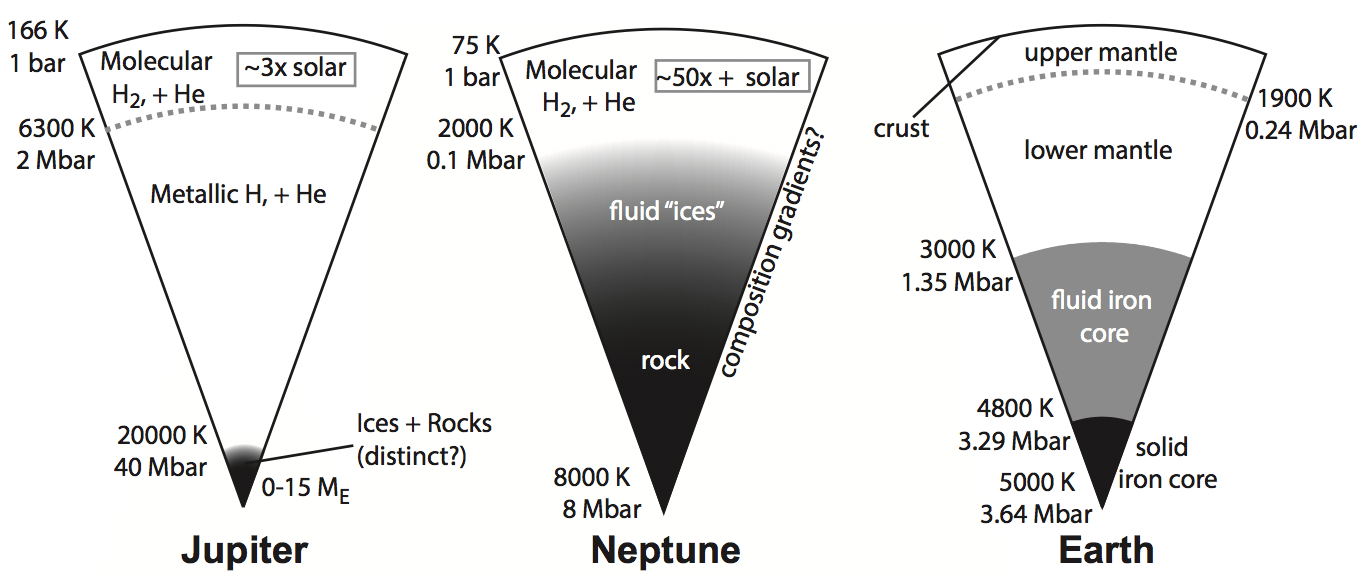}
\end{center}
\caption{\small Pie-slice diagrams of a representative gas-giant
  (Jupiter), ice-giant (Neptune), and terrestrial (Earth) planets.
  Jupiter's outer $\sim$20\% in radius consists mostly of molecular
  hydrogen.  Pressures and temperatures quickly increase inside the
  planet and the bulk of Jupiter is made of liquid metallic hydrogen,
  a form of liquid hydrogen that is highly electrically conductive.
  At its center there may be a concentration of heavier elements, but
  constraints are not firm.  The visible atmosphere, and perhaps the
  entire H/He envelope is enhanced in metals by a factor of 3-5
  compared to the Sun.  Within Neptune, most or perhaps all of the
  hydrogen is found in molecular form, while the bulk of the planet's
  mass, the ``middle'' gray layer, is composed of fluid conducting
  molecules and molecular fragments probably consisting of H$_2$O,
  NH$_3$, and CH$_4$.  Constrains on the interior rock-to-ice ratio
  are not firm, and it is not clear how distinct the layer boundaries
  actually are.  The visible atmosphere is enhanced in carbon by a
  factor of $\sim$50 compared to the Sun, but there is little
  constraint on other elements.  Within the Earth the mantle is
  composed of silicates whose mineralogy differs between the upper and
  the lower mantle.  The distinct core is mostly iron-nickle, with
  small admixtures of unknown lighter elements.  The solid core is
  currently growing at the expense of the liquid core. Most of the
  interiors of all of the planets are thought to be convective.}
\label{fig:JupNepEarth}
\end{figure*}

Giant planets where most of the planet's mass is composed of heavy
elements, rather than mostly H/He gas, begin our transition to our
next class of planets.  These so-called ``Neptune-class'' planets
still have a thick H/He envelope, but the light-element envelope does
not comprise the majority of the planet's mass.  In our Solar System,
Uranus (14.5~$M_\oplus$ and 4.0~\re), and Neptune (17.1~$M_\oplus$ and
3.9~\re) are our examples of these planets.
Figure~\ref{fig:JupNepEarth} portrays how the bulk structure of a
Neptune-class planet differs from that of either a Jupiter or an Earth
(which we will address in the next section).

Uranus and Neptune are generally known for their bluish color and are
often lumped together as two ``ice giants'' because most structure
models find that the majority of the planetary mass is in a deep fluid
ionic sea probably consisting predominantly of water, and also
containing ammonia and methane.  While the planets appear outwardly
very similar, there is ample evidence that the planet's interiors are
quite different.  The diversity within our two Neptune-class planets
should be a clear reminder that this class of exoplanets should harbor
tremendous diversity.

First, both planets do \emph{not} simply have homogeneous three-layer
structures with an H/He upper envelope, water-dominated middle
envelope, and rocky core.  Neither planet is as centrally condensed as
this often-suggested but too-simple picture would imply.  Uranus is
more centrally condensed that Neptune.  More dramatically it also has
a heat flux from its deep interior that is no more than 10\% that of
Neptune, which may be due to deep composition gradients that suppress
large-scale convection.  Uranus is also flipped over on its side with
its spin axis nearly in its orbital plane, which might imply that the
stochastic nature of giant collisions near the end of the
planet-formation era plays a major role in determining the structure
of this class of planet.  Even today, much of our knowledge of the
structure and evolution of these planets remains uncertain and
provisional \cite{fortney_et_al2011, podolak+helled2012,
  nettelmann_et_al2013}.

Both planets have H/He atmospheres that are strongly enriched in
metals.  Only the carbon abundance (in methane) can be measuredly
fairly definitively via spectroscopy, and requires a carbon
enhancement of about 50$\times$ solar in each planet
\cite{fletcher_et_al2010}.  Indirect evidence suggests that oxygen may
be several hundred times solar in Neptune \cite{luszcz+de_pater2013}.
This suggests a framework where the relatively thin H/He envelopes of
Neptune-class planets may well be extremely enhanced in metals
compared to their parent stars.

Whether Neptune-class exoplanets are true ``ice giants'' (meaning that
much of their mass is made up of water and other fluid ``planetary
ices'') is an open question.  If planets form beyond the ``ice line''
where water has condensed (like all of our giant planets), then a
substantial fraction of the mass of Neptune-class planets is probably
water.  However, planets may also form within the ice line, in which
case they would have rock/iron interiors with an accreted gaseous
envelope.

The first Neptune-class exoplanet with a measured mass and radius was
the transiting ``hot Neptune'' GJ 436b, which is 22.2~\me\ and
4.3~$R_\oplus$ \cite{butler_et_al2004, gillon_et_al2007}.  The
planet's location on the mass/radius diagram implies that the planet
must have a substantial H/He envelope and is not composed only of
water, for instance.  The relative mass fractions of H/He, water (and
other icy components like ammonia and methane), rock, and iron cannot
be ascertained from mass and radius alone.  Even with the assumption
of a fixed rock:iron ratio, one can mix a wide array of 3-component
compositions of rock/iron, water, and H/He.  For instance, one could
model the planet with only H/He and rock, or with only H/He and water,
ignoring a third component.  Another complication is that one needs to
be able to model the thermal evolution of the planet to understand the
deep interior temperatures and densities of the possible components of
this several-Gyr-old planet.  Degeneracy in inferred composition for
Neptune class exoplanets will always be the rule
\cite{adams_et_al2008, nettelmann_et_al2010}.

Diversity in exo-Neptune interior structure can be seen from a
comparison of GJ 436b to Kepler-30d.  Models of GJ 436b generally
suggest an interior that is 80-90\% heavy elements (similar to Uranus
and Neptune).  However, Kepler-30d has nearly the same mass (23.1~\me)
but a radius twice as large (8.8~\re) \cite{sanchis_et_al2012}.  This
suggests that the planet is only 30\% heavy elements, and 70\% H/He
gas, thereby already breaking our ``rule'' that planets of this class
(mass-range) should not be made predominantly of hydrogen!

{\it Kepler} data show a dramatic increase in planetary occurrence
from 6~\re\ to 2~\re, indicating that ``sub-Neptunes'' from 2-3~\re\
are a very common planetary type \cite{fressin_et_al2013,
  dong+zhu2013}.  Such planets also generally need a H/He envelope,
but one that comprises perhaps only $\sim$1-5\% of the planetary mass.
Several of these objects have masses in the 3-10~\me\ range,
indicating that even relatively small planets can accrete and maintain
gaseous H/He envelopes, even despite ongoing evaporative mass loss
\cite{lopez_et_al2012}.

While the majority of {\it Kepler} planets are around distant, faint
stars, there is also a relatively nearby example of this class of
sub-Neptune planet, named GJ~1214b.  This $\sim$500-K transiting
planet orbits its cool star in a close-in orbit, and its mass
(6.5~\me) and radius (2.7~\re) are well determined
\cite{charbonneau_et_al2009} and seem to imply a gaseous component
atop a liquid/solid core \cite{rogers+seager2010b, kipping_et_al2013}.
A series of observational campaigns have attempted to characterize the
visible H/He atmosphere of the planet.  This would perhaps allow the
composition of the entire H/He envelope to be constrained.  However,
the observed transit-radius spectrum has been nearly featureless
\cite{bean_et_al2010, berta_et_al2012}.  This suggests either that
cloud material is obscuring the atmosphere or that the atmosphere is
quite compact, so that it imprints little signal on the stellar
transmitted light.  If the planet's atmosphere is strongly enriched
with metals, at a level even higher than Uranus and Neptune, this
could greatly increase the mean molecular mass and reduce the
atmosphere's vertical height, although such a high atmospheric
mean-molecular weight is disfavored by the observed mass and radius
\cite{kipping_et_al2013}.  Once the atmospheres of more of these
planets have been probed in more detail, we can begin to make firmer
connections that could link the composition of the H/He envelope with
planetary mass and orbital location.

Below some mass ($\lesssim$3--5~\me?), cores are small enough that
they either do not accrete nebular H/He gas, or the gas that is
accreted is quickly lost.  Such objects are true terrestrial or water
planets, often called super-Earths.  We now turn our attention to
these larger cousins of Earth and Venus.

\section{Terrestrial and Ocean Planets}
\label{sec:terrestrialocean}
The search for ``Earth-like,'' terrestrial planets is a major
objective in the study of exoplanets.  Terrestrial planets are objects
such as Mars, Venus, and Mercury that are composed predominantly of
elements such as Si, Mg, Fe, O, and, perhaps in some cases, C
\cite{sotin_et_al2007}.  Now that exoplanets have been discovered that
are roughly the size of Venus or Earth, one key question is whether a
given terrestrial exoplanet is more similar to Venus or to Earth, as
discussed in \S{\bf Terrestrial Planets} below.  However, other types
of planets or satellites may be habitable and, in some sense,
Earth-like.  In the Solar System, Enceladus and Europa share with the
Earth the rare characteristic of possessing an ocean that is in
contact with a rocky interior.  It is not known whether there is life
on Enceladus or Europa, but their existence motivates the study of
exoplanets that would have a large fraction of H$_2$O, known as ocean
planets \cite{leger_et_al2004}.  As discussed in \S{\bf Ocean
  Planets}, it might be difficult to determine unambiguously that an
exoplanet actually is an ocean planet, because such worlds could
occupy the same region of the $M_p$/$R_p$ plane as planets consisting
of rocky cores surrounded by atmospheres of H/He
\cite{seager+deming2010}.

\subsection{Terrestrial Planets}
\label{ssec:terrestrial}
Venus and Earth have about the same mass and radius, and Venus's
slightly smaller density could be explained by its smaller mass.  That
is, the uncompressed density is the same for the two planets.  But
Venus and Earth have evolved to present-day conditions so different as
to invite the question of whether distance to the Sun is the only
parameter that drives such a different evolution.

\begin{figure*}[t]
\begin{center}
\includegraphics[width=14cm,angle=0,clip=true]{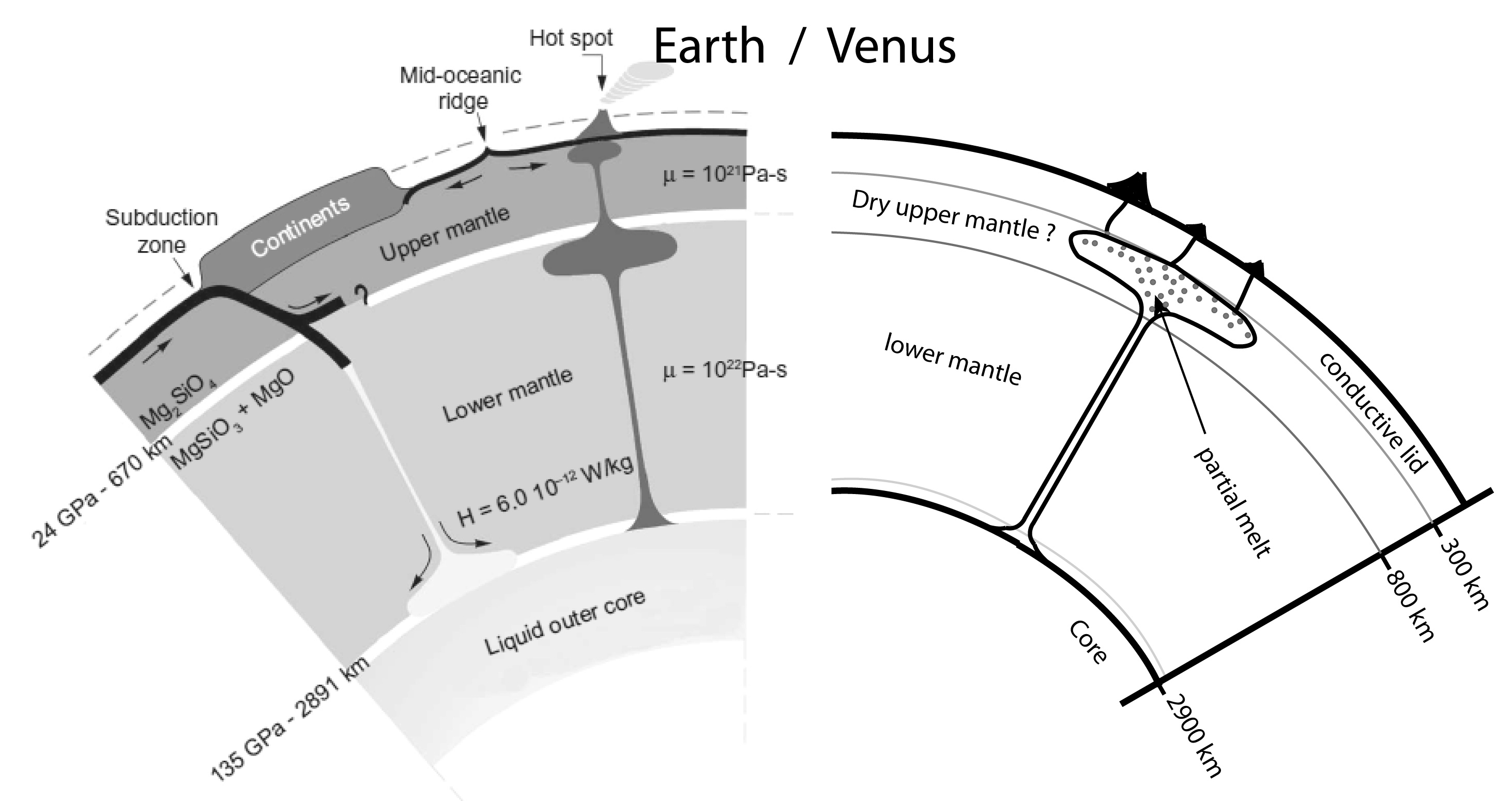}
\end{center}
\caption{\small Schematic representation of the interior structure and
  dynamics of the Earth (left) and Venus (right).  On both planets,
  hot plumes form at the core/mantle boundary, although the heat that
  comes out of core is a fraction ($\sim$25\% for the Earth) of the
  total heat released at the surface.  The mineralogical
  transformation at the upper/lower mantle interface may or may not be
  a barrier to convection: the left panel depicts a hot plume that
  would be stopped at this interface; whereas the right panel shows
  how the plume moves up to the bottom of the conductive lid. The
  depth of this interface on Venus is larger than that of the Earth
  because the gravity on Venus is lower. On Earth (left), the oceanic
  lithosphere (black) forms at the mid-ocean ridges and is recycled in
  the mantle at subduction zones (adapted from \cite{sotin_et_al2011,
    smrekar+sotin2012}).}
\label{fig:EarthVenus}
\end{figure*}

Among the major differences between Venus and Earth is the lack of
current plate tectonics on Venus, as revealed by the geological
analysis of radar images acquired by the Magellan mission
\cite{schaber_et_al1992, basilevsky+head2002}.  Plate tectonics plays
an important role on Earth, providing a very efficient way of cooling
the Earth's interior.  It also provides an exchange mechanism between
the interior, the surface, and the atmosphere: at subduction zones,
hydrated minerals and sediments are carried back to the mantle; at
mid-ocean ridges, new crust and volatiles contained in the mantle are
released into the atmosphere (Fig.~\ref{fig:EarthVenus}).  Whether
such a cycle \cite{walker_et_al1981} is required for life to form
\cite{spiegel+turner2012} and evolve on a terrestrial planet is
debated.  Still, since the only world where life has been identified
has plate tectonics, it is crucial to understand what controls
tectonic dynamics.

Plate tectonics occurs in response to convective motions in the mantle
(Fig.~\ref{fig:EarthVenus}).  The main features are the formation of
instabilities at the thermal boundaries located at the core-mantle
transition and at the surface (Fig.~\ref{fig:EarthVenus}).  Hot plumes
are common features of Venus and Earth
(e.g. \cite{smrekar+sotin2012}).  Venus is in the so-called
stagnant-lid regime --- cold plumes form at a cold thermal-boundary
layer, deep below the surface, at the base of the ``stagnant lid''
through which heat is transferred by conduction.  The stagnant lid
regime is not efficient in removing heat from the interior
\cite{reese_et_al1998}.  The tectonic regime (plate tectonics or
stagnant lid) depends on a number of parameters, including the size of
the planet, the surface temperature, the presence of water at the
surface, and more generally the history of the planet
\cite{lenardic+crowley2012}. The reason why Venus and Earth have
evolved so differently pathways remains uncertain
\cite{stevenson2003}.

Convective stresses scale with the vigor of convection.  Some authors
argue that larger planets have larger convective stresses and,
therefore, are more likely to experience plate tectonics
\cite{valencia_et_al2007, valencia_et_al2009}.  However, the yield
strength depends on more than just the size of the planet.  Recent
work has shown that the depth dependence of the crustal yield
strength, a poorly constrained parameter, is critical in determining
the convective regime and the likelihood of plate tectonics
\cite{lenardic+crowley2012}: the more the yield strength increases
with depth, the lower the probability of plate tectonics.  The yield
strength and its variation with depth depend on surface temperature,
the presence of liquid water, the geological history of the planet
(such as whether large impact craters have weakened the lithosphere),
and the presence of light, Earth-like continents
(Fig.~\ref{fig:EarthVenus}), at whose border the lithosphere is
weaker.  Other geological properties, such as the temperature
dependence of viscosity, the amount of interior heating, and
mineralogical transformations in the mantle, also influence the
propensity for plate tectonics.  Furthermore, episodic regimes in
which stagnant-lid periods alternate with active-lid periods may also
exist \cite{lenardic+crowley2012}.  One hypothesis that has been
proposed to explain Venus's global resurfacing is a transition from a
plate-tectonics regime to a stagnant-lid regime about 1~Gyr ago
\cite{stevenson2003}.

Definitively determining whether a planet around another star
undergoes plate tectonics will be extremely difficult.  Still, as our
understanding of geology on Earth and elsewhere in the Solar System
improves, we might be better able to estimate the likelihood of
exoplanetary plate tectonics.

\subsection{Ocean Planets}
\label{ssec:ocean}
Water is a key ingredient for the formation and development of
life.\footnote{It is also possible that having too much water might
  hinder the development of life, because it might both dilute
  important nutrients and disrupt geochemical thermal regulation
  processes.}  ``Follow the water'' has, therefore, been a motto for
Mars exploration.  Although H$_2$O is known to exist in ice and vapor
form on Mars, subsurface liquid water has not yet been identified.  On
the other hand, geophysical observations (magnetic field and gravity
field) by the Galileo mission and the Cassini mission strongly suggest
the presence of water under the icy crust of Callisto, Ganymede,
Europa, Titan, and Enceladus.  This has inspired several studies about
the possibility of ocean-dominated exoplanets.

Ocean exoplanets are not known to exist, but some of their possible
properties have been theoretically explored.  For a given mass, a
planet that is 50\% (by mass) H$_2$O and 50\% Earth-like composition
would have a radius $\sim$25\% larger than that of a terrestrial
exoplanet \cite{leger_et_al2004}.  As illustrated in
Fig.~\ref{fig:Rad_vs_stuff}, several exoplanets have ($M_p,R_p$)
values that lie in the vicinity of the $R_p[M_p]$ relationship for
Earth-like composition.  Recent work has examined the structure of
ocean planets with a variety of water fractions and found,
generically, that ocean planets have $R_p[M_p]$ relations that are
somewhat larger at a given mass than Earth-like planets
\cite{grasset_et_al2009, rogers+seager2010a, zeng+sasselov2013,
  kipping_et_al2013}, with the degree of increased radius depending on
the water fraction.  However, there are degeneracies: for instance, a
planet whose ($M_p,R_p$) location is consistent with being an ocean
planet could also have a silicate (terrestrial) core veiled by a H/He
rich atmosphere \cite{seager+deming2010}.  This degeneracy can be
resolved if the atmospheric composition can be discerned
\cite{miller-ricci_et_al2009a}.

An ocean planet's habitability could be affected by whether its liquid
water is in contact with a rocky core.  Planets with large amounts of
H$_2$O might develop a high-pressure ice layer between the core and
the liquid layer \cite{grasset_et_al2009}, similar to the structure
proposed for Ganymede, Callisto, and Titan.  Counterintuitively, more
massive planets might have thinner oceans, because they have greater
pressure gradients and, therefore, their oceans more quickly enter the
high-pressure ice-layer regime.  In order to have contact between the
ocean and the rocky core, the fraction of H$_2$O has to be small, as
is the case on Earth ($2\times 10^{-4}$) and Europa ($7\times
10^{-2}$).  However, in some circumstances, even if the planet has a
large water fraction, there can still be contact between a fluid water
layer and a silicate layer.  This would occur if the temperature at
the top of the H$_2$O layer is high enough that the envelope's
temperature gradient yields a temperature at the base of the water
layer that is that is still in the fluid regime, and might be a
natural outcome in Neptune-class planets with or without the loss of
their primordial H$_2$-He atmospheres.

\section{Conclusions}
\label{sec:conc}
Observational campaigns in the next decade will help us to refine our
knowledge of the structure of planets both in our Solar System and
beyond.  The Juno Mission \cite{bolton2010} will provide crucial new
insights into Jupiter's internal structure, and should help resolve
the longstanding question of how much water is in Jupiter.  Beyond the
Solar System, the recently approved {\it Transiting Exoplanet Survey
  Satellite} (\emph{TESS}) will identify many planets around stars
close enough and bright enough that the planets are amenable to
follow-up observations from the ground or with the {\it James Webb
  Space Telescope} (\emph{JWST}).  Learning about both the masses (via
RV measurements) and about the atmospheres of these planets will
inform our understanding of their bulk compositions and structures.

Finally, in some cases, giant exoplanets that are sufficiently far
from their stars and sufficiently self-luminous, such as those in the
HR~8799 system \cite{marois_et_al2008}, may be directly imaged with
new, high-contrast imaging techniques.  These direct observations of
young objects are sensitive to planets' initial conditions and,
therefore, might help us \pn{i} to distinguish between formation
mechanisms of widely-separated, young jovian objects
\cite{marley_et_al2007, fortney_et_al2008b, spiegel+burrows2012}, and
\pn{ii} to learn about their interior structures.  The ground-based
programs that will undertake such surveys include the Gemini Planet
Imager (GPI, \cite{macintosh_et_al2006}), the Spectro-Polarimetric
High-contrast Exoplanet REsearch instrument (SPHERE) on the Very Large
Telescope \cite{beuzit_et_al2008}, and more.  Exoplanetary
observations have revealed unanticipated structures in planets both
large and small, and so the prospect of a flood of upcoming data
promises more surprises and new insight into comparative planetology.

\begin{acknowledgments}
  DSS gratefully acknowledges support from NSF grant AST-0807444, the
  Keck Fellowship, the Friends of the Institute, and the AMIAS.  JF
  acknowledges NSF grant AST-1010017.  CS acknowledges support by NAI
  Icy Worlds.
\end{acknowledgments}


\begin{thebibliography}{10}
\expandafter\ifx\csname url\endcsname\relax
  \def\url#1{\texttt{#1}}\fi
\expandafter\ifx\csname urlprefix\endcsname\relax\def\urlprefix{URL }\fi
\providecommand{\bibinfo}[2]{#2}
\providecommand{\eprint}[2][]{\url{#2}}

\bibitem{borucki_et_al2011}
\bibinfo{author}{{Borucki}, W.~J.} \emph{et~al.}
\newblock \bibinfo{title}{{Characteristics of Planetary Candidates Observed by
  Kepler. II. Analysis of the First Four Months of Data}}.
\newblock \emph{\bibinfo{journal}{\apj}} \textbf{\bibinfo{volume}{736}},
  \bibinfo{pages}{19} (\bibinfo{year}{2011}).
\newblock \eprint{1102.0541}.

\bibitem{batalha_et_al2013}
\bibinfo{author}{{Batalha}, N.~M.} \emph{et~al.}
\newblock \bibinfo{title}{{Planetary Candidates Observed by Kepler. III.
  Analysis of the First 16 Months of Data}}.
\newblock \emph{\bibinfo{journal}{\apjs}} \textbf{\bibinfo{volume}{204}},
  \bibinfo{pages}{24} (\bibinfo{year}{2013}).
\newblock \eprint{1202.5852}.

\bibitem{morton+johnson2011}
\bibinfo{author}{{Morton}, T.~D.} \& \bibinfo{author}{{Johnson}, J.~A.}
\newblock \bibinfo{title}{{On the Low False Positive Probabilities of Kepler
  Planet Candidates}}.
\newblock \emph{\bibinfo{journal}{\apj}} \textbf{\bibinfo{volume}{738}},
  \bibinfo{pages}{170} (\bibinfo{year}{2011}).
\newblock \eprint{1101.5630}.

\bibitem{fressin_et_al2013}
\bibinfo{author}{{Fressin}, F.} \emph{et~al.}
\newblock \bibinfo{title}{{The False Positive Rate of Kepler and the Occurrence
  of Planets}}.
\newblock \emph{\bibinfo{journal}{\apj}} \textbf{\bibinfo{volume}{766}},
  \bibinfo{pages}{81} (\bibinfo{year}{2013}).
\newblock \eprint{1301.0842}.

\bibitem{bakos_et_al2004}
\bibinfo{author}{{Bakos}, G.} \emph{et~al.}
\newblock \bibinfo{title}{{Wide-Field Millimagnitude Photometry with the HAT: A
  Tool for Extrasolar Planet Detection}}.
\newblock \emph{\bibinfo{journal}{\pasp}} \textbf{\bibinfo{volume}{116}},
  \bibinfo{pages}{266--277} (\bibinfo{year}{2004}).
\newblock \eprint{arXiv:astro-ph/0401219}.

\bibitem{pollacco_et_al2006}
\bibinfo{author}{{Pollacco}, D.~L.} \emph{et~al.}
\newblock \bibinfo{title}{{The WASP Project and the SuperWASP Cameras}}.
\newblock \emph{\bibinfo{journal}{PASP}} \textbf{\bibinfo{volume}{118}},
  \bibinfo{pages}{1407--1418} (\bibinfo{year}{2006})
\newblock \eprint{arXiv:astro-ph/0608454}

\bibitem{wright_et_al2011}
\bibinfo{author}{{Wright}, J.~T.} \emph{et~al.}
\newblock \bibinfo{title}{{The Exoplanet Orbit Database}}.
\newblock \emph{\bibinfo{journal}{PASP}} \textbf{\bibinfo{volume}{123}},
  \bibinfo{pages}{412--422} (\bibinfo{year}{2011}).
\newblock \eprint{1012.5676}.

\bibitem{schneider_et_al2011}
\bibinfo{author}{{Schneider}, J.}, \bibinfo{author}{{Dedieu}, C.},
  \bibinfo{author}{{Le Sidaner}, P.}, \bibinfo{author}{{Savalle}, R.} \&
  \bibinfo{author}{{Zolotukhin}, I.}
\newblock \bibinfo{title}{{Defining and cataloging exoplanets: the exoplanet.eu
  database}}.
\newblock \emph{\bibinfo{journal}{A \& A}} \textbf{\bibinfo{volume}{532}},
  \bibinfo{pages}{A79} (\bibinfo{year}{2011})
\newblock \eprint{1106.0586}

\bibitem{zapolsky+salpeter1969}
\bibinfo{author}{{Zapolsky}, H.~S.} \& \bibinfo{author}{{Salpeter}, E.~E.}
\newblock \bibinfo{title}{{The Mass-Radius Relation for Cold Spheres of Low
  Mass}}.
\newblock \emph{\bibinfo{journal}{\apj}} \textbf{\bibinfo{volume}{158}},
  \bibinfo{pages}{809--+} (\bibinfo{year}{1969}).

\bibitem{burrows_et_al2001}
\bibinfo{author}{{Burrows}, A.}, \bibinfo{author}{{Hubbard}, W.~B.},
  \bibinfo{author}{{Lunine}, J.~I.} \& \bibinfo{author}{{Liebert}, J.}
\newblock \bibinfo{title}{{The theory of brown dwarfs and extrasolar giant
  planets}}.
\newblock \emph{\bibinfo{journal}{Reviews of Modern Physics}}
  \textbf{\bibinfo{volume}{73}}, \bibinfo{pages}{719--765}
  (\bibinfo{year}{2001}).
\newblock \eprint{arXiv:astro-ph/0103383}.

\bibitem{guillot_et_al1996}
\bibinfo{author}{{Guillot}, T.}, \bibinfo{author}{{Burrows}, A.},
  \bibinfo{author}{{Hubbard}, W.~B.}, \bibinfo{author}{{Lunine}, J.~I.} \&
  \bibinfo{author}{{Saumon}, D.}
\newblock \bibinfo{title}{{Giant Planets at Small Orbital Distances}}.
\newblock \emph{\bibinfo{journal}{\apjl}} \textbf{\bibinfo{volume}{459}},
  \bibinfo{pages}{L35+} (\bibinfo{year}{1996}).
\newblock \eprint{astro-ph/9511109}.

\bibitem{saumon_et_al1996}
\bibinfo{author}{{Saumon}, D.} \emph{et~al.}
\newblock \bibinfo{title}{{A Theory of Extrasolar Giant Planets}}.
\newblock \emph{\bibinfo{journal}{ApJ}} \textbf{\bibinfo{volume}{460}},
  \bibinfo{pages}{993--+} (\bibinfo{year}{1996}).
\newblock \eprint{arXiv:astro-ph/9510046}.

\bibitem{charbonneau_et_al2000}
\bibinfo{author}{{Charbonneau}, D.}, \bibinfo{author}{{Brown}, T.~M.},
  \bibinfo{author}{{Latham}, D.~W.} \& \bibinfo{author}{{Mayor}, M.}
\newblock \bibinfo{title}{{Detection of Planetary Transits Across a Sun-like
  Star}}.
\newblock \emph{\bibinfo{journal}{\apjl}} \textbf{\bibinfo{volume}{529}},
  \bibinfo{pages}{L45--L48} (\bibinfo{year}{2000}).
\newblock \eprint{astro-ph/9911436}.

\bibitem{burrows_et_al2000}
\bibinfo{author}{{Burrows}, A.} \emph{et~al.}
\newblock \bibinfo{title}{{On the Radii of Close-in Giant Planets}}.
\newblock \emph{\bibinfo{journal}{\apjl}} \textbf{\bibinfo{volume}{534}},
  \bibinfo{pages}{L97--L100} (\bibinfo{year}{2000}).
\newblock \eprint{astro-ph/0003185}.

\bibitem{baraffe_et_al2010}
\bibinfo{author}{{Baraffe}, I.}, \bibinfo{author}{{Chabrier}, G.} \&
  \bibinfo{author}{{Barman}, T.}
\newblock \bibinfo{title}{{The physical properties of extra-solar planets}}.
\newblock \emph{\bibinfo{journal}{Reports on Progress in Physics}}
  \textbf{\bibinfo{volume}{73}}, \bibinfo{pages}{016901}
  (\bibinfo{year}{2010}).
\newblock \eprint{1001.3577}.

\bibitem{fortney+nettelmann2010}
\bibinfo{author}{{Fortney}, J.~J.} \& \bibinfo{author}{{Nettelmann}, N.}
\newblock \bibinfo{title}{{The Interior Structure, Composition, and Evolution
  of Giant Planets}}.
\newblock \emph{\bibinfo{journal}{Space Science Reviews}}
  \textbf{\bibinfo{volume}{152}}, \bibinfo{pages}{423--447}
  (\bibinfo{year}{2010})
\newblock \eprint{0912.0533}.

\bibitem{saumon_et_al1995}
\bibinfo{author}{{Saumon}, D.}, \bibinfo{author}{{Chabrier}, G.} \&
  \bibinfo{author}{{van Horn}, H.~M.}
\newblock \bibinfo{title}{{An Equation of State for Low-Mass Stars and Giant
  Planets}}.
\newblock \emph{\bibinfo{journal}{\apjs}} \textbf{\bibinfo{volume}{99}},
  \bibinfo{pages}{713} (\bibinfo{year}{1995}).

\bibitem{militzer+hubbard2013}
\bibinfo{author}{{Militzer}, B.} \& \bibinfo{author}{{Hubbard}, W.~B.}
\newblock \bibinfo{title}{{Ab Initio Equation of State for Hydrogen-Helium
  Mixtures with Recalibration of the Giant-planet M-R Relation}}.
\newblock \emph{\bibinfo{journal}{ApJ}} \textbf{\bibinfo{volume}{774}},
  \bibinfo{pages}{148} (\bibinfo{year}{2013})
\newblock \eprint{1302.4691}

\bibitem{guillot+showman2002}
\bibinfo{author}{{Guillot}, T.} \& \bibinfo{author}{{Showman}, A.~P.}
\newblock \bibinfo{title}{{Evolution of ``51 Pegasus b-like'' planets}}.
\newblock \emph{\bibinfo{journal}{\aap}} \textbf{\bibinfo{volume}{385}},
  \bibinfo{pages}{156--165} (\bibinfo{year}{2002}).
\newblock \eprint{astro-ph/0202234}.

\bibitem{burrows_et_al2007}
\bibinfo{author}{{Burrows}, A.}, \bibinfo{author}{{Hubeny}, I.},
  \bibinfo{author}{{Budaj}, J.} \& \bibinfo{author}{{Hubbard}, W.~B.}
\newblock \bibinfo{title}{{Possible Solutions to the Radius Anomalies of
  Transiting Giant Planets}}.
\newblock \emph{\bibinfo{journal}{\apj}} \textbf{\bibinfo{volume}{661}},
  \bibinfo{pages}{502--514} (\bibinfo{year}{2007}).
\newblock \eprint{arXiv:astro-ph/0612703}.

\bibitem{perna_et_al2010b}
\bibinfo{author}{{Perna}, R.}, \bibinfo{author}{{Menou}, K.} \&
  \bibinfo{author}{{Rauscher}, E.}
\newblock \bibinfo{title}{{Ohmic Dissipation in the Atmospheres of Hot
  Jupiters}}.
\newblock \emph{\bibinfo{journal}{\apj}} \textbf{\bibinfo{volume}{724}},
  \bibinfo{pages}{313--317} (\bibinfo{year}{2010}).
\newblock \eprint{1009.3273}.

\bibitem{bodenheimer_et_al2001}
\bibinfo{author}{{Bodenheimer}, P.}, \bibinfo{author}{{Lin}, D.~N.~C.} \&
  \bibinfo{author}{{Mardling}, R.~A.}
\newblock \bibinfo{title}{{On the Tidal Inflation of Short-Period Extrasolar
  Planets}}.
\newblock \emph{\bibinfo{journal}{\apj}} \textbf{\bibinfo{volume}{548}},
  \bibinfo{pages}{466--472} (\bibinfo{year}{2001}).

\bibitem{miller_et_al2009}
\bibinfo{author}{{Miller}, N.}, \bibinfo{author}{{Fortney}, J.~J.} \&
  \bibinfo{author}{{Jackson}, B.}
\newblock \bibinfo{title}{{Inflating and Deflating Hot Jupiters: Coupled Tidal
  and Thermal Evolution of Known Transiting Planets}}.
\newblock \emph{\bibinfo{journal}{ApJ}} \textbf{\bibinfo{volume}{702}},
  \bibinfo{pages}{1413--1427} (\bibinfo{year}{2009})
\newblock \eprint{0907.1268}

\bibitem{batygin+stevenson2010}
\bibinfo{author}{{Batygin}, K.} \& \bibinfo{author}{{Stevenson}, D.~J.}
\newblock \bibinfo{title}{{Inflating Hot Jupiters with Ohmic Dissipation}}.
\newblock \emph{\bibinfo{journal}{\apjl}} \textbf{\bibinfo{volume}{714}},
  \bibinfo{pages}{L238--L243} (\bibinfo{year}{2010}).
\newblock \eprint{1002.3650}.

\bibitem{wu+lithwick2013}
\bibinfo{author}{{Wu}, Y.} \& \bibinfo{author}{{Lithwick}, Y.}
\newblock \bibinfo{title}{{Ohmic Heating Suspends, Not Reverses, the Cooling
  Contraction of Hot Jupiters}}.
\newblock \emph{\bibinfo{journal}{\apj}} \textbf{\bibinfo{volume}{763}},
  \bibinfo{pages}{13} (\bibinfo{year}{2013}).
\newblock \eprint{1202.0026}.

\bibitem{grasset_et_al2009}
\bibinfo{author}{{Grasset}, O.}, \bibinfo{author}{{Schneider}, J.} \&
  \bibinfo{author}{{Sotin}, C.}
\newblock \bibinfo{title}{{A Study of the Accuracy of M-R Relationships
  for Si-Rich and H2O-Rich Planets up to 100 Earth Masses}}.
\newblock \emph{\bibinfo{journal}{ApJ}} \textbf{\bibinfo{volume}{693}},
  \bibinfo{pages}{722--733} (\bibinfo{year}{2009})
\newblock \eprint{0902.1640}

\bibitem{marois_et_al2008}
\bibinfo{author}{{Marois}, C.} \emph{et~al.}
\newblock \bibinfo{title}{{Direct Imaging of Multiple Planets Orbiting the Star
  HR 8799}}.
\newblock \emph{\bibinfo{journal}{Science}} \textbf{\bibinfo{volume}{322}},
  \bibinfo{pages}{1348--} (\bibinfo{year}{2008}).
\newblock \eprint{0811.2606}.

\bibitem{arras+socrates2010}
\bibinfo{author}{{Arras}, P.} \& \bibinfo{author}{{Socrates}, A.}
\newblock \bibinfo{title}{{Thermal Tides in Fluid Extrasolar Planets}}.
\newblock \emph{\bibinfo{journal}{ApJ}} \textbf{\bibinfo{volume}{714}},
  \bibinfo{pages}{1--12} (\bibinfo{year}{2010})
\newblock \eprint{0912.2313}

\bibitem{budaj_et_al2012}
\bibinfo{author}{{Budaj}, J.}, \bibinfo{author}{{Hubeny}, I.} \&
  \bibinfo{author}{{Burrows}, A.}
\newblock \bibinfo{title}{{Day and night side core cooling of a strongly
  irradiated giant planet}}.
\newblock \emph{\bibinfo{journal}{\aap}} \textbf{\bibinfo{volume}{537}},
  \bibinfo{pages}{A115} (\bibinfo{year}{2012}).
\newblock \eprint{1111.5478}.

\bibitem{spiegel+burrows2013}
\bibinfo{author}{{Spiegel}, D.~S.} \& \bibinfo{author}{{Burrows}, A.}
\newblock \bibinfo{title}{{Thermal Processes Governing Hot-Jupiter Radii}}.
\newblock \emph{\bibinfo{journal}{\apj}} \textbf{\bibinfo{volume}{772}},
  \bibinfo{pages}{76} (\bibinfo{year}{2013}).
\newblock \eprint{1303.0293}.

\bibitem{baraffe_et_al2003}
\bibinfo{author}{{Baraffe}, I.}, \bibinfo{author}{{Chabrier}, G.},
  \bibinfo{author}{{Barman}, T.~S.}, \bibinfo{author}{{Allard}, F.} \&
  \bibinfo{author}{{Hauschildt}, P.~H.}
\newblock \bibinfo{title}{{Evolutionary models for cool brown dwarfs and
  extrasolar giant planets. The case of HD 209458}}.
\newblock \emph{\bibinfo{journal}{\aap}} \textbf{\bibinfo{volume}{402}},
  \bibinfo{pages}{701--712} (\bibinfo{year}{2003}).
\newblock \eprint{arXiv:astro-ph/0302293}.

\bibitem{fortney_et_al2011b}
\bibinfo{author}{{Fortney}, J.~J.} \emph{et~al.}
\newblock \bibinfo{title}{{Discovery and Atmospheric Characterization of Giant
  Planet Kepler-12b: An Inflated Radius Outlier}}.
\newblock \emph{\bibinfo{journal}{\apjs}} \textbf{\bibinfo{volume}{197}},
  \bibinfo{pages}{9} (\bibinfo{year}{2011}).
\newblock \eprint{1109.1611}.

\bibitem{weiss_et_al2013}
\bibinfo{author}{{Weiss}, L.~M.} \emph{et~al.}
\newblock \bibinfo{title}{{The Mass of KOI-94d and a Relation for Planet
  Radius, Mass, and Incident Flux}}.
\newblock \emph{\bibinfo{journal}{\apj}} \textbf{\bibinfo{volume}{768}},
  \bibinfo{pages}{14} (\bibinfo{year}{2013}).
\newblock \eprint{1303.2150}.

\bibitem{mizuno_et_al1978}
\bibinfo{author}{{Mizuno}, H.}, \bibinfo{author}{{Nakazawa}, K.} \&
  \bibinfo{author}{{Hayashi}, C.}
\newblock \bibinfo{title}{{Instability of a gaseous envelope surrounding a
  planetary core and formation of giant planets}}.
\newblock \emph{\bibinfo{journal}{Progress of Theoretical Physics}}
  \textbf{\bibinfo{volume}{60}}, \bibinfo{pages}{699--710}
  (\bibinfo{year}{1978}).

\bibitem{boss2010}
\bibinfo{author}{{Boss}, A.~P.}
\newblock \bibinfo{title}{{Giant Planet Formation by Disk Instability in Low
  Mass Disks?}}
\newblock \emph{\bibinfo{journal}{\apjl}} \textbf{\bibinfo{volume}{725}},
  \bibinfo{pages}{L145--L149} (\bibinfo{year}{2010}).
\newblock \eprint{1010.5819}.

\bibitem{guillot_et_al2006}
\bibinfo{author}{{Guillot}, T.} \emph{et~al.}
\newblock \bibinfo{title}{{A correlation between the heavy element content of
  transiting extrasolar planets and the metallicity of their parent stars}}.
\newblock \emph{\bibinfo{journal}{A \& A}} \textbf{\bibinfo{volume}{453}},
  \bibinfo{pages}{L21--L24} (\bibinfo{year}{2006})
\newblock \eprint{arXiv:astro-ph/0605751}

\bibitem{sato_et_al2005}
\bibinfo{author}{{Sato}, B.} \emph{et~al.}
\newblock \bibinfo{title}{{The N2K Consortium. II. A Transiting Hot Saturn
  around HD 149026 with a Large Dense Core}}.
\newblock \emph{\bibinfo{journal}{\apj}} \textbf{\bibinfo{volume}{633}},
  \bibinfo{pages}{465--473} (\bibinfo{year}{2005}).
\newblock \eprint{astro-ph/0507009}.

\bibitem{fortney_et_al2006b}
\bibinfo{author}{{Fortney}, J.~J.}, \bibinfo{author}{{Saumon}, D.},
  \bibinfo{author}{{Marley}, M.~S.}, \bibinfo{author}{{Lodders}, K.} \&
  \bibinfo{author}{{Freedman}, R.~S.}
\newblock \bibinfo{title}{{Atmosphere, Interior, and Evolution of the
  Metal-rich Transiting Planet HD 149026b}}.
\newblock \emph{\bibinfo{journal}{\apj}} \textbf{\bibinfo{volume}{642}},
  \bibinfo{pages}{495--504} (\bibinfo{year}{2006}).
\newblock \eprint{arXiv:astro-ph/0507422}.

\bibitem{fortney_et_al2011}
\bibinfo{author}{{Fortney}, J.~J.}, \bibinfo{author}{{Ikoma}, M.},
  \bibinfo{author}{{Nettelmann}, N.}, \bibinfo{author}{{Guillot}, T.} \&
  \bibinfo{author}{{Marley}, M.~S.}
\newblock \bibinfo{title}{{Self-consistent Model Atmospheres and the Cooling of
  the Solar System's Giant Planets}}.
\newblock \emph{\bibinfo{journal}{\apj}} \textbf{\bibinfo{volume}{729}},
  \bibinfo{pages}{32--+} (\bibinfo{year}{2011})
\newblock \eprint{1101.0606}

\bibitem{podolak+helled2012}
\bibinfo{author}{{Podolak}, M.} \& \bibinfo{author}{{Helled}, R.}
\newblock \bibinfo{title}{{What Do We Really Know about Uranus and Neptune?}}
\newblock \emph{\bibinfo{journal}{\apjl}} \textbf{\bibinfo{volume}{759}},
  \bibinfo{pages}{L32} (\bibinfo{year}{2012}).
\newblock \eprint{1208.5551}.

\bibitem{nettelmann_et_al2013}
\bibinfo{author}{{Nettelmann}, N.}, \bibinfo{author}{{Helled}, R.},
  \bibinfo{author}{{Fortney}, J.~J.} \& \bibinfo{author}{{Redmer}, R.}
\newblock \bibinfo{title}{{New indication for a dichotomy in the interior
  structure of Uranus and Neptune from the application of modified shape and
  rotation data}}.
\newblock \emph{\bibinfo{journal}{Planetary and Space Science}}
  \textbf{\bibinfo{volume}{77}}, \bibinfo{pages}{143--151}
  (\bibinfo{year}{2013})
\newblock \eprint{1207.2309}

\bibitem{fletcher_et_al2010}
\bibinfo{author}{{Fletcher}, L.~N.}, \bibinfo{author}{{Drossart}, P.},
  \bibinfo{author}{{Burgdorf}, M.}, \bibinfo{author}{{Orton}, G.~S.} \&
  \bibinfo{author}{{Encrenaz}, T.}
\newblock \bibinfo{title}{{Neptune's atmospheric composition from AKARI
  infrared spectroscopy}}.
\newblock \emph{\bibinfo{journal}{A \& A}} \textbf{\bibinfo{volume}{514}},
  \bibinfo{pages}{A17} (\bibinfo{year}{2010})
\newblock \eprint{1003.5571}

\bibitem{luszcz+de_pater2013}
\bibinfo{author}{{Luszcz-Cook}, S.~H.} \& \bibinfo{author}{{de Pater}, I.}
\newblock \bibinfo{title}{{Constraining the origins of Neptune's carbon
  monoxide abundance with CARMA millimeter-wave observations}}.
\newblock \emph{\bibinfo{journal}{\icarus}} \textbf{\bibinfo{volume}{222}},
  \bibinfo{pages}{379--400} (\bibinfo{year}{2013}).
\newblock \eprint{1301.1990}.

\bibitem{butler_et_al2004}
\bibinfo{author}{{Butler}, R.~P.} \emph{et~al.}
\newblock \bibinfo{title}{{A Neptune-Mass Planet Orbiting the Nearby M Dwarf GJ
  436}}.
\newblock \emph{\bibinfo{journal}{\apj}} \textbf{\bibinfo{volume}{617}},
  \bibinfo{pages}{580--588} (\bibinfo{year}{2004}).
\newblock \eprint{arXiv:astro-ph/0408587}.

\bibitem{gillon_et_al2007}
\bibinfo{author}{{Gillon}, M.} \emph{et~al.}
\newblock \bibinfo{title}{{Accurate Spitzer infrared radius measurement for the
  hot Neptune GJ 436b}}.
\newblock \emph{\bibinfo{journal}{\aap}} \textbf{\bibinfo{volume}{471}},
  \bibinfo{pages}{L51--L54} (\bibinfo{year}{2007}).
\newblock \eprint{arXiv:0707.2261}.

\bibitem{adams_et_al2008}
\bibinfo{author}{{Adams}, E.~R.}, \bibinfo{author}{{Seager}, S.} \&
  \bibinfo{author}{{Elkins-Tanton}, L.}
\newblock \bibinfo{title}{{Ocean Planet or Thick Atmosphere: On the Mass-Radius
  Relationship for Solid Exoplanets with Massive Atmospheres}}.
\newblock \emph{\bibinfo{journal}{\apj}} \textbf{\bibinfo{volume}{673}},
  \bibinfo{pages}{1160--1164} (\bibinfo{year}{2008}).
\newblock \eprint{arXiv:0710.4941}.

\bibitem{nettelmann_et_al2010}
\bibinfo{author}{{Nettelmann}, N.}, \bibinfo{author}{{Kramm}, U.},
  \bibinfo{author}{{Redmer}, R.} \& \bibinfo{author}{{Neuh{\"a}user}, R.}
\newblock \bibinfo{title}{{Interior structure models of GJ 436b}}.
\newblock \emph{\bibinfo{journal}{\aap}} \textbf{\bibinfo{volume}{523}},
  \bibinfo{pages}{A26+} (\bibinfo{year}{2010}).
\newblock \eprint{1002.4447}.

\bibitem{sanchis_et_al2012}
\bibinfo{author}{{Sanchis-Ojeda}, R.} \emph{et~al.}
\newblock \bibinfo{title}{{Alignment of the stellar spin with the orbits of a
  three-planet system}}.
\newblock \emph{\bibinfo{journal}{\nat}} \textbf{\bibinfo{volume}{487}},
  \bibinfo{pages}{449--453} (\bibinfo{year}{2012}).
\newblock \eprint{1207.5804}.

\bibitem{dong+zhu2013}
\bibinfo{author}{{Dong}, S.} \& \bibinfo{author}{{Zhu}, Z.}
\newblock \bibinfo{title}{{Fast Rise of Neptune-size Planets from 10 to 250
  Days -- Statistics of Kepler Planet Candidates up to 0.75 AU}}.
\newblock \emph{\bibinfo{journal}{\apj}} \textbf{\bibinfo{volume}{778}},
  \bibinfo{pages}{53} (\bibinfo{year}{2013}).
\newblock \eprint{1212.4853}.

\bibitem{lopez_et_al2012}
\bibinfo{author}{{Lopez}, E.~D.}, \bibinfo{author}{{Fortney}, J.~J.} \&
  \bibinfo{author}{{Miller}, N.}
\newblock \bibinfo{title}{{How Thermal Evolution and Mass-loss Sculpt
  Populations of Super-Earths and Sub-Neptunes: Application to the Kepler-11
  System and Beyond}}.
\newblock \emph{\bibinfo{journal}{\apj}} \textbf{\bibinfo{volume}{761}},
  \bibinfo{pages}{59} (\bibinfo{year}{2012}).
\newblock \eprint{1205.0010}.

\bibitem{charbonneau_et_al2009}
\bibinfo{author}{{Charbonneau}, D.} \emph{et~al.}
\newblock \bibinfo{title}{{A super-Earth transiting a nearby low-mass star}}.
\newblock \emph{\bibinfo{journal}{\nat}} \textbf{\bibinfo{volume}{462}},
  \bibinfo{pages}{891--894} (\bibinfo{year}{2009}).
\newblock \eprint{0912.3229}.

\bibitem{rogers+seager2010b}
\bibinfo{author}{{Rogers}, L.~A.} \& \bibinfo{author}{{Seager}, S.}
\newblock \bibinfo{title}{{Three Possible Origins for the Gas Layer on GJ
  1214b}}.
\newblock \emph{\bibinfo{journal}{\apj}} \textbf{\bibinfo{volume}{716}},
  \bibinfo{pages}{1208--1216} (\bibinfo{year}{2010}).
\newblock \eprint{0912.3243}.

\bibitem{kipping_et_al2013}
\bibinfo{author}{{Kipping}, D.~M.}, \bibinfo{author}{{Spiegel}, D.~S.} \&
  \bibinfo{author}{{Sasselov}, D.~D.}
\newblock \bibinfo{title}{{A simple, quantitative method to infer the minimum
  atmospheric height of small exoplanets}}.
\newblock \emph{\bibinfo{journal}{\mnras}} \textbf{\bibinfo{volume}{434}},
  \bibinfo{pages}{1883--1888} (\bibinfo{year}{2013}).
\newblock \eprint{1306.3221}.

\bibitem{bean_et_al2010}
\bibinfo{author}{{Bean}, J.~L.}, \bibinfo{author}{{Miller-Ricci Kempton}, E.}
  \& \bibinfo{author}{{Homeier}, D.}
\newblock \bibinfo{title}{{A ground-based transmission spectrum of the
  super-Earth exoplanet GJ 1214b}}.
\newblock \emph{\bibinfo{journal}{\nat}} \textbf{\bibinfo{volume}{468}},
  \bibinfo{pages}{669--672} (\bibinfo{year}{2010}).
\newblock \eprint{1012.0331}.

\bibitem{berta_et_al2012}
\bibinfo{author}{{Berta}, Z.~K.} \emph{et~al.}
\newblock \bibinfo{title}{{The Flat Transmission Spectrum of the Super-Earth
  GJ1214b from Wide Field Camera 3 on the Hubble Space Telescope}}.
\newblock \emph{\bibinfo{journal}{\apj}} \textbf{\bibinfo{volume}{747}},
  \bibinfo{pages}{35} (\bibinfo{year}{2012})
\newblock \eprint{1111.5621}

\bibitem{sotin_et_al2007}
\bibinfo{author}{{Sotin}, C.}, \bibinfo{author}{{Grasset}, O.} \&
  \bibinfo{author}{{Mocquet}, A.}
\newblock \bibinfo{title}{{Mass radius curve for extrasolar Earth-like planets
  and ocean planets}}.
\newblock \emph{\bibinfo{journal}{\icarus}} \textbf{\bibinfo{volume}{191}},
  \bibinfo{pages}{337--351} (\bibinfo{year}{2007}).

\bibitem{leger_et_al2004}
\bibinfo{author}{{L{\'e}ger}, A.} \emph{et~al.}
\newblock \bibinfo{title}{{A new family of planets? ``Ocean-Planets''}}.
\newblock \emph{\bibinfo{journal}{\icarus}} \textbf{\bibinfo{volume}{169}},
  \bibinfo{pages}{499--504} (\bibinfo{year}{2004})
\newblock \eprint{arXiv:astro-ph/0308324}

\bibitem{seager+deming2010}
\bibinfo{author}{{Seager}, S.} \& \bibinfo{author}{{Deming}, D.}
\newblock \bibinfo{title}{{Exoplanet Atmospheres}}.
\newblock \emph{\bibinfo{journal}{ARAA}} \textbf{\bibinfo{volume}{48}},
  \bibinfo{pages}{631--672} (\bibinfo{year}{2010})
\newblock \eprint{1005.4037}

\bibitem{schaber_et_al1992}
\bibinfo{author}{{Schaber}, G.~G.} \emph{et~al.}
\newblock \bibinfo{title}{{Geology and distribution of impact craters on Venus
  - What are they telling us?}}
\newblock \emph{\bibinfo{journal}{Journal of Geophysical Research}}
  \textbf{\bibinfo{volume}{97}}, \bibinfo{pages}{13257} (\bibinfo{year}{1992}).

\bibitem{basilevsky+head2002}
\bibinfo{author}{{Basilevsky}, A.~T.} \& \bibinfo{author}{{Head}, J.~W.}
\newblock \bibinfo{title}{{Venus: Timing and rates of geologic activity}}.
\newblock \emph{\bibinfo{journal}{Geology}} \textbf{\bibinfo{volume}{30}},
  \bibinfo{pages}{1015} (\bibinfo{year}{2002}).

\bibitem{walker_et_al1981}
\bibinfo{author}{{Walker}, J.~C.~G.}, \bibinfo{author}{{Hays}, P.~B.} \&
  \bibinfo{author}{{Kasting}, J.~F.}
\newblock \bibinfo{title}{{A Negative Feedback Mechanism For The Long-Term
  Stabilization of Earth's Surface Temperature}}.
\newblock \emph{\bibinfo{journal}{Journal of Geophysical Research}}
  \textbf{\bibinfo{volume}{86}}, \bibinfo{pages}{9776--9782}
  (\bibinfo{year}{1981}).

\bibitem{spiegel+turner2012}
\bibinfo{author}{{Spiegel}, D.~S.} \& \bibinfo{author}{{Turner}, E.~L.}
\newblock \bibinfo{title}{{Bayesian analysis of the astrobiological
  implications of life's early emergence on Earth}}.
\newblock \emph{\bibinfo{journal}{Proceedings of the National Academy of
  Science}} \textbf{\bibinfo{volume}{109}}, \bibinfo{pages}{395--400}
  (\bibinfo{year}{2012}).
\newblock \eprint{1107.3835}.

\bibitem{smrekar+sotin2012}
\bibinfo{author}{{Smrekar}, S.~E.} \& \bibinfo{author}{{Sotin}, C.}
\newblock \bibinfo{title}{{Constraints on mantle plumes on Venus: Implications
  for volatile history}}.
\newblock \emph{\bibinfo{journal}{\icarus}} \textbf{\bibinfo{volume}{217}},
  \bibinfo{pages}{510--523} (\bibinfo{year}{2012}).

\bibitem{reese_et_al1998}
\bibinfo{author}{{Reese}, C.~C.}, \bibinfo{author}{{Solomatov}, V.~S.} \&
  \bibinfo{author}{{Moresi}, L.-N.}
\newblock \bibinfo{title}{{Heat transport efficiency for stagnant lid
  convection with dislocation viscosity: Application to Mars and Venus}}.
\newblock \emph{\bibinfo{journal}{Journal of Geophysical Research}}
  \textbf{\bibinfo{volume}{103}}, \bibinfo{pages}{13643--13658}
  (\bibinfo{year}{1998}).

\bibitem{lenardic+crowley2012}
\bibinfo{author}{{Lenardic}, A.} \& \bibinfo{author}{{Crowley}, J.~W.}
\newblock \bibinfo{title}{{On the Notion of Well-defined Tectonic Regimes for
  Terrestrial Planets in this Solar System and Others}}.
\newblock \emph{\bibinfo{journal}{\apj}} \textbf{\bibinfo{volume}{755}},
  \bibinfo{pages}{132} (\bibinfo{year}{2012}).

\bibitem{stevenson2003}
\bibinfo{author}{{Stevenson}, D.}
\newblock \bibinfo{title}{{Les styles de convection du manteau et leur
  influence sur l'{\'e}volution des plan{\`e}tes}}.
\newblock \emph{\bibinfo{journal}{Comptes Rendus Geoscience}}
  \textbf{\bibinfo{volume}{335}}, \bibinfo{pages}{99--111}
  (\bibinfo{year}{2003}).

\bibitem{valencia_et_al2007}
\bibinfo{author}{{Valencia}, D.}, \bibinfo{author}{{Sasselov}, D.~D.} \&
  \bibinfo{author}{{O'Connell}, R.~J.}
\newblock \bibinfo{title}{{Detailed Models of Super-Earths: How Well Can We
  Infer Bulk Properties?}}
\newblock \emph{\bibinfo{journal}{\apj}} \textbf{\bibinfo{volume}{665}},
  \bibinfo{pages}{1413--1420} (\bibinfo{year}{2007})
\newblock \eprint{arXiv:0704.3454}

\bibitem{valencia_et_al2009}
\bibinfo{author}{{Valencia}, D.}, \bibinfo{author}{{Ikoma}, M.},
  \bibinfo{author}{{Guillot}, T.} \& \bibinfo{author}{{Nettelmann}, N.}
\newblock \bibinfo{title}{{Composition and fate of short-period super-Earths.
  The case of CoRoT-7b}}.
\newblock \emph{\bibinfo{journal}{\aap}} \textbf{\bibinfo{volume}{516}},
  \bibinfo{pages}{A20} (\bibinfo{year}{2010})
\newblock \eprint{0907.3067}

\bibitem{rogers+seager2010a}
\bibinfo{author}{{Rogers}, L.~A.} \& \bibinfo{author}{{Seager}, S.}
\newblock \bibinfo{title}{{A Framework for Quantifying the Degeneracies of
  Exoplanet Interior Compositions}}.
\newblock \emph{\bibinfo{journal}{\apj}} \textbf{\bibinfo{volume}{712}},
  \bibinfo{pages}{974--991} (\bibinfo{year}{2010}).
\newblock \eprint{0912.3288}.

\bibitem{zeng+sasselov2013}
\bibinfo{author}{{Zeng}, L.} \& \bibinfo{author}{{Sasselov}, D.}
\newblock \bibinfo{title}{{A Detailed Model Grid for Solid Planets from 0.1
  through 100 Earth Masses}}.
\newblock \emph{\bibinfo{journal}{\pasp}} \textbf{\bibinfo{volume}{125}},
  \bibinfo{pages}{227--239} (\bibinfo{year}{2013}).
\newblock \eprint{1301.0818}.

\bibitem{miller-ricci_et_al2009a}
\bibinfo{author}{{Miller-Ricci}, E.}, \bibinfo{author}{{Seager}, S.} \&
  \bibinfo{author}{{Sasselov}, D.}
\newblock \bibinfo{title}{{The Atmospheric Signatures of Super-Earths: How to
  Distinguish Between Hydrogen-Rich and Hydrogen-Poor Atmospheres}}.
\newblock \emph{\bibinfo{journal}{\apj}} \textbf{\bibinfo{volume}{690}},
  \bibinfo{pages}{1056--1067} (\bibinfo{year}{2009}).
\newblock \eprint{0808.1902}.

\bibitem{sotin_et_al2011}
\bibinfo{author}{{Sotin}, C.}, \bibinfo{author}{{Jackson}, J.~M.} \&
  \bibinfo{author}{{Seager}, S.}
\newblock \emph{\bibinfo{title}{{Terrestrial Planet Interiors}}},
  \bibinfo{pages}{375--395} (\bibinfo{year}{2011}).

\bibitem{bolton2010}
\bibinfo{author}{{Bolton}, S.~J.} \& \bibinfo{author}{{Bolton}}.
\newblock \bibinfo{title}{{The Juno Mission}}.
\newblock In \emph{\bibinfo{booktitle}{IAU Symposium}}, vol.
  \bibinfo{volume}{269} of \emph{\bibinfo{series}{IAU Symposium}},
  \bibinfo{pages}{92--100} (\bibinfo{year}{2010}).

\bibitem{marley_et_al2007}
\bibinfo{author}{{Marley}, M.~S.}, \bibinfo{author}{{Fortney}, J.~J.},
  \bibinfo{author}{{Hubickyj}, O.}, \bibinfo{author}{{Bodenheimer}, P.} \&
  \bibinfo{author}{{Lissauer}, J.~J.}
\newblock \bibinfo{title}{{On the Luminosity of Young Jupiters}}.
\newblock \emph{\bibinfo{journal}{\apj}} \textbf{\bibinfo{volume}{655}},
  \bibinfo{pages}{541--549} (\bibinfo{year}{2007}).
\newblock \eprint{arXiv:astro-ph/0609739}.

\bibitem{fortney_et_al2008b}
\bibinfo{author}{{Fortney}, J.~J.}, \bibinfo{author}{{Marley}, M.~S.},
  \bibinfo{author}{{Saumon}, D.} \& \bibinfo{author}{{Lodders}, K.}
\newblock \bibinfo{title}{{Synthetic Spectra and Colors of Young Giant Planet
  Atmospheres: Effects of Initial Conditions and Atmospheric Metallicity}}.
\newblock \emph{\bibinfo{journal}{\apj}} \textbf{\bibinfo{volume}{683}},
  \bibinfo{pages}{1104--1116} (\bibinfo{year}{2008}).
\newblock \eprint{0805.1066}.

\bibitem{spiegel+burrows2012}
\bibinfo{author}{{Spiegel}, D.~S.} \& \bibinfo{author}{{Burrows}, A.}
\newblock \bibinfo{title}{{Spectral and Photometric Diagnostics of Giant Planet
  Formation Scenarios}}.
\newblock \emph{\bibinfo{journal}{\apj}} \textbf{\bibinfo{volume}{745}},
  \bibinfo{pages}{174} (\bibinfo{year}{2012}).
\newblock \eprint{1108.5172}.

\bibitem{macintosh_et_al2006}
\bibinfo{author}{{Macintosh}, B.} \emph{et~al.}
\newblock \bibinfo{title}{{The Gemini Planet Imager}}.
\newblock In \emph{\bibinfo{booktitle}{Society of Photo-Optical Instrumentation
  Engineers (SPIE) Conference Series}}, vol. \bibinfo{volume}{6272} of
  \emph{\bibinfo{series}{Society of Photo-Optical Instrumentation Engineers
  (SPIE) Conference Series}} (\bibinfo{year}{2006}).

\bibitem{beuzit_et_al2008}
\bibinfo{author}{{Beuzit}, J.-L.} \emph{et~al.}
\newblock \bibinfo{title}{{SPHERE: a planet finder instrument for the VLT}}.
\newblock In \emph{\bibinfo{booktitle}{Society of Photo-Optical Instrumentation
  Engineers (SPIE) Conference Series}}, vol. \bibinfo{volume}{7014} of
  \emph{\bibinfo{series}{Society of Photo-Optical Instrumentation Engineers
  (SPIE) Conference Series}} (\bibinfo{year}{2008}).

\end{thebibliography}

\end{article}

\end{document}